\documentclass{emulateapj}
\newcommand{\second}{\ensuremath{\mathrm{s}}}
\newcommand{\hour}{\ensuremath{\mathrm{h}}}
\newcommand{\kmps}{\ensuremath{\mathrm{km\ s}^{-1}}}

\newcommand{\feh}{\ensuremath{\mathrm{[Fe/H]}}}
\newcommand{\snr}{\ensuremath{\mathrm{SNR}}}
\newcommand{\logg}{\ensuremath{\log{g}}}
\newcommand{\teff}{\ensuremath{T_{\mathrm{eff}}}}
\newcommand{\rv}{\ensuremath{\mathrm{RV}}}
\newcommand{\rvs}{\ensuremath{\mathrm{RVs}}}
\newcommand{\mad}{\ensuremath{\mathrm{MAD}}}
\newcommand{\median}{\ensuremath{\mathrm{median}}}
\providecommand{\e}[1]{\ensuremath{\times 10^{#1}}}

\newcommand{\logp}{\ensuremath{\log P}}
\newcommand{\loga}{\ensuremath{\log A}}
\newcommand{\vnaught}{\ensuremath{V_0}}

\newcommand{\fehlimitA}{\ensuremath{-1.43}}
\newcommand{\fehlimitB}{\ensuremath{-0.66}}

\newcommand{\eoi}{\ensuremath{e/i}}
\newcommand{\err}{\ensuremath{\sigma_\mathrm{RV}}}

\newcommand{\bmodel}{\ensuremath{\mathcal{M}_b}}
\newcommand{\smodel}{\ensuremath{\mathcal{M}_s}}
\newcommand{\pixres}{\ensuremath{70\ \mathrm{km\ s}^{-1}\ \mathrm{pixel}^{-1}}}

\newcommand{\fbin}{\ensuremath{f_b}}

\shorttitle{Statistical Time-Resolved Spectroscopy}
\shortauthors{Hettinger et al.}

\begin{document}

\title{statistical time-resolved spectroscopy: a higher fraction of short-period binaries for metal-rich F-type dwarfs in SDSS}

\author{T. Hettinger}
\affil{Department of Physics and Astronomy, Michigan State University,
  East Lansing, MI 48824, USA}
\email{hettin12@msu.edu}

\author{C. Badenes}
\affil{Department of Physics and Astronomy and Pittsburgh Particle Physics, Astrophysics and Cosmology Center (PITT PACC), University of Pittsburgh, Pittsburgh, PA 15260, USA}
\email{badenes@pitt.edu}

\author{J. Strader}
\affil{Department of Physics and Astronomy, Michigan State University,
  East Lansing, MI 48824, USA}
\email{strader@pa.msu.edu}

\author{S.J. Bickerton}
\affil{Kavli Institute for the Physics and Mathematics of the Universe (WPI), Todai Institutes for Advanced Study, the University of Tokyo, Chiba, 277-8583, Japan}
\email{steven.bickerton@ipmu.jp}

\and

\author{T.C. Beers}
\affil{Dept. of Physics and JINA Center for the Evolution of the Elements, Univ. of Notre Dame, Notre Dame, IN 46556, USA}
\email{tbeers@nd.edu}

\begin{abstract}
  Stellar multiplicity lies at the heart of many problems in modern astrophysics, including the physics of star formation, the observational properties of unresolved stellar populations, and the rates of interacting binaries such as cataclysmic variables, X-ray binaries, and Type Ia supernovae. However, little is known about the stellar multiplicity of field stars in the Milky Way, in particular about the differences in the multiplicity characteristics between metal-rich disk stars and metal-poor halo stars. In this study we perform a statistical analysis of $\sim$14,000 F-type dwarf stars in the Milky Way through time-resolved spectroscopy with the sub-exposures archived in the Sloan Digital Sky Survey. We obtain absolute radial velocity measurements through template cross-correlation of individual sub-exposures with temporal baselines varying from minutes to years. These sparsely sampled radial velocity curves are analyzed using Markov chain Monte Carlo techniques to constrain the very short-period binary fraction for field F-type stars in the Milky Way. Metal-rich disk stars were found to be $30\%$ more likely to have companions with periods shorter than 12 days than metal-poor halo stars.
\end{abstract}

\section{Introduction}
  \label{intro}
  Stellar multiplicity plays a crucial role in many fields of astronomy.  Star formation and evolution, Galactic chemical evolution, nuclear astrophysics, and cosmology are all influenced by our understanding of the multiplicity properties of an underlying stellar population.  Binary interactions lead to phenomena as diverse as cataclysmic variables, classical novae, X-ray binaries, gamma-ray bursts, and Type Ia supernovae.  Stellar interactions are also the cause of the anomalous surface abundances measured in Ba stars, CH stars, and the majority of carbon-enhanced metal-poor stars \citep{lucatello05}.  The rates of these phenomena depend on the multiplicity properties such as the fraction of stars with companions and the distributions of separations and mass ratios.  How these properties are in turn affected by variables such as stellar age, metallicity, and dynamical environment remains poorly understood.  \cite{moe13} find no significant trends with metallicity for O- and B-stars, but more work is needed for lower-mass stars.

  The recent review by \cite{duchene13} summarizes the state of the art in multiplicity studies.  The fraction of systems with companions is known to be a strong function of stellar mass \citep{lada06, raghavan10, clark12}, and there are hints that lower mass systems have smaller separations \citep{duquennoy91, allen07, raghavan10}.  Studies of the Solar neighborhood also indicate that lower metallicity stars are more likely to have stellar companions \citep{raghavan10}.

  These results are based on heterogeneous samples of a few hundred stars at most, often dominated by wide systems which will never become interacting binaries.  The spectroscopic surveys that reach small periods are labor intensive because large numbers of radial velocities (\rvs) are required to find the orbital solution of each target.  This leads to small sample sizes, which have only increased modestly in the past two decades, from 167 in \cite{duquennoy91} to 454 in \cite{raghavan10}.  The drive to collect complete samples has limited previous spectroscopic studies to the Solar neighborhood or specific stellar clusters, but neither of these strategies can probe the full range of metallicities and ages spanning the field stars of the Milky Way (MW) disk and halo components.  These limits bias the interpretation of data against the global properties of, and variation within, the MW field.  Thus, we are motivated to take a statistical approach with a sample of stars located throughout the field in order to investigate their multiplicity properties with respect to age, \feh, and component membership.

  With the advent of multiplexed spectroscopic surveys like SDSS \citep{york00} and LAMOST \citep{cui12}, we can use multiple \rv\ measurements of thousands of stars to study the properties of stellar multiplicity that are more representative of the entire Galaxy.  SDSS Data Release 8 \citep{aihara11} contains over 1.8 million optical spectra from the original SDSS spectrographs including over 600,000 stellar spectra.  In this work we employ a lesser known SDSS feature, the time-resolved dimension.  To facilitate cosmic ray removal, spectra were constructed through co-addition of several individual sub-exposures, typically 15 minutes in duration.  Although under-utilized, the benefit of the sub-exposure domain is recognized in works such as \cite{badenes09} and \cite{bickerton12}.  Portions of the sky were also re-observed for calibration and scientific purposes.  These additional pointings, combined with the sub-exposures, yield a time dimension where single stars have exposure coverage ranging from 3 sub-exposures up to over 40 sub-exposures, and time gaps from hours to nearly a decade.  The techniques employed herein follow the time-resolved work by \cite{badenes12} and \cite{maoz12}.

\section{Measurements}
  \label{sec:method}

  \subsection{SDSS Observations and Sample Selection}
    \label{ssec:sample}
    F-type dwarfs are chosen for our sample because of the large number of stars targeted by SDSS with repeat observations, and their relatively mild variability and activity.  Additionally, F-stars have main sequence (MS) lifetimes greater than $5\ \mathrm{Gyr}$, allowing us to select MS stars from both the younger disk and older halo.  The Sloan Stellar Parameter Pipeline (SSPP; \citealp{lee08}) was developed to determine parameters for stellar spectra in the SDSS archive, including metallicity \feh, effective temperature \teff, and surface gravity \logg.  Sample selection began with identifying science primary objects from SEGUE-1 \citep{yanny09} and SEGUE-2 (Rockosi et al., in prep.) in the SSPP that were classified as an F-type star by the ``Hammer'' classification code \citep{covey07}.  To minimize the effects of stellar evolution on multiplicity, we selected only dwarf stars ($\logg\geq 3.75$).  Stars with multiple fiber pluggings were identified astrometrically and joined with the appropriate science primary fibers.

    After measuring stellar \rvs\ (Section \ref{ssec:rv}), systematics were revealed in the SDSS sub-exposure spectra.  These correlations appear as similar shifts in \rvs\ for many fibers located on the same plate, typically affecting neighboring fibers on the CCD.  After plate-wide comparisons of F-stars, \rv\ correlations were corrected where possible.  Corrections applied to the $10^4$ \rvs\ are as large as $17\ \kmps$ with a standard deviation of $2.2\ \kmps$.  Not all correlations could be identified automatically because of multiple groups of correlated shifts, opposite in direction, on some plates.  Visual inspection of plates containing numerous false binary detections lead to the removal of 25 plates including 1155 stars.  We urge individuals using sub-exposure spectroscopy in SDSS to consider these systematic shifts in the wavelength solutions.

\begin{figure*}[htbp]
      \centering
      \plotone{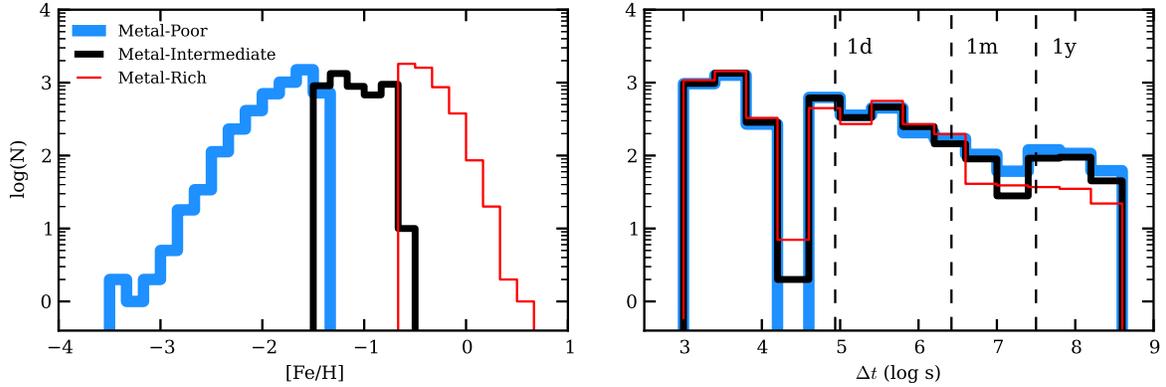}
      \caption{Left: Metallicity distribution for 14,302 F-dwarfs. Right: Distribution of maximum time lag between the first and last exposure of a star.}
      \label{fig:feh_baseline}
    \end{figure*}

    Quality control consisted of the removal of: stars without valid parameters in SSPP, fibers located on `bad' plates, sub-exposures with a median pixel signal-to-noise ratio (\snr) less than 20 or with fewer than 3000 unflagged pixels, stars with time lags $\Delta T<1800\ \second$, stars with less than three clean sub-exposures, and corrupt or misclassified spectra (from visual inspection of stars with the largest RV variation or non-characteristic \teff).  The final sample consists of 14,302 stars (16,894 fibers) with as many as 47 sub-exposures, spanning up to nine years of observations (Figure \ref{fig:feh_baseline}).

    Our cleaned sample is characterized by metallicities ranging from $-3.41\leq\feh\leq+0.52$.  To aid comparison in our analysis, the final sample was sub-divided into three groups of equal size by cuts in metallicity at $\feh=\fehlimitA$ and $\feh=\fehlimitB$ (Figure \ref{fig:feh_baseline}).  The majority of the stars have three or four sub-exposures ($\mathrm{median} = 4$), typically taken about 15 minutes apart.  The median time lag for a star is 2 hours, however more than three years between observations can be seen in more than 250 stars (Figure \ref{fig:feh_baseline}).  \snr s for sub-exposures lie in the range $20<\snr<84$ with a median value of $32$.  

  \subsection{Radial Velocities}
    \label{ssec:rv}
    \rv\ measurements were attained through cross-correlation of sub-exposures with a master template constructed from 7207 sample-star, co-added spectra where the co-added $\snr > 50$.  The spectra were de-shifted using the redshift value assigned to the co-adds by the SDSS pipeline, continuum-normalized, and averaged together.

\begin{figure*}[htbp]
      \centering
      \plotone{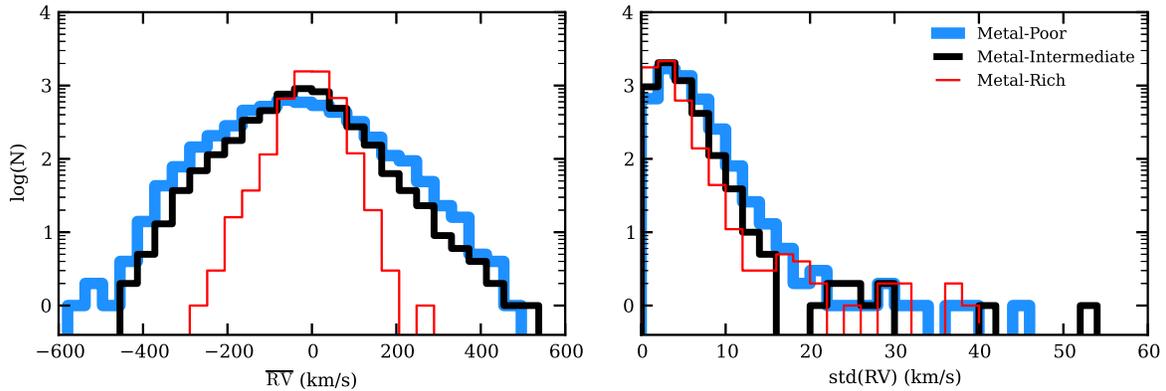}
      \caption{Mean (left) and standard deviation (right) of radial velocities within a star.  Variations in the standard deviation of velocities are affected, in part, by the larger measurement uncertainties for metal-poorer stars.}
      \label{fig:comboRV}
    \end{figure*}

    Sub-exposures were independently prepared and cross-correlated with the template.  Spectra were continuum-normalized by dividing the spectrum with a highly smoothed version of itself using a FFT smoothing algorithm, and then cross-correlated with the template at various integer pixel lags.  Each spectrum had a cross-correlation function (CCF) that was fit with a smooth spline interpolation.  With spectral resolution of $R\sim2000$, the peak lag in pixels translates to the spectrum's redshift at \pixres.  The mean and standard deviation of \rvs\ for individual stars are shown in the Figure \ref{fig:comboRV} distributions.  The velocity dispersion of the mean RVs decreases with increasing \feh, indicating that our \feh-groups sample both the disk and halo components of the MW.  The standard deviation of \rvs\ within individual stars is larger for the metal-poor group; however, empirically estimated uncertainties also show larger measurement errors for metal-poor stars.  This underscores the importance of the use of proper error analysis in a method such as ours.

  \subsection{Uncertainties}
    \label{ssec:uncertainties}
    It is well known that uncertainties in CCF peaks must be estimated empirically or through some Monte Carlo method (e.g.,~\citealt{peterson98}).  For this work we determined \rv\ uncertainties empirically by quantifying the spread in measurements for spectra of similar quality.  The median absolute deviation (\mad) is a robust measure of the variability of a sample and is related to the standard deviation by $\sigma = 1.4826\mad$, where $\mad = \median(\left|\rv_i - \median(\rv)\right|)$ \citep{leys13}.  All measurements were de-shifted into the rest frame using the SDSS estimates of the co-add redshift, and placed into bins of similar metallicity ($\feh\pm 0.25$) and signal-to-noise ($\snr\pm 2.5$).  Initial tests showed no correlations between measurement spreads and either \logg~or \teff.  Estimates for the uncertainty of \rv\ measurements within a bin were calculated using \mad\ values.  Here, it is assumed that the majority of stars do not have detectable variability over the observed time baseline, and that effects from intrinsic variations in \rv\ are minimized by adopting median values.  After performing this process for all bins, a functional form for assigning \rv\ measurement uncertainties \err\ was fit with an inverse proportionality to \snr, and with a linear correction in \feh.  The measurement uncertainty as a function of \feh\ and \snr\ is, in \kmps,
    \begin{equation}
      \label{eq:empiricalerror}
      \err\left(\feh,\snr\right) = \frac{(-26.51\feh + 50.52)}{\snr} + 1.23.
    \end{equation}
    Uncertainties are sub-pixel, falling below the spectral resolution of \pixres.  For exposures with $\snr<25$, uncertainties range from $3.0$ to $8.0\ \kmps$, with a median value of $5.0\ \kmps$.  Exposures with $\snr>40$ have uncertainties in the range $1.9$ to $4.4\ \kmps$, with a median value of $2.7\ \kmps$.

\section{Multiplicity}
  \label{sec:multiplicity}
  The probability of a star having a companion was determined through model comparison using a trans-dimensional, hierarchical, Markov chain Monte Carlo (MCMC) method.  Two models were compared: a single-star model \smodel, and a binary-star model \bmodel.  The hyperparameter $\lambda$, indexes the model choice at each step in the MCMC chain.  We evaluated the hierarchical model using the Python package \textit{emcee}, a MCMC ensemble sampler \citep{foreman13}.
  
  The single-star model \smodel, fits a star with non-varying \rvs, parameterized by a systemic velocity \vnaught.  Because intra-plate systematics are known to exist, it is reasonable to assume inter-plate systematics exist as well.  In light of this, $(P-1)$ additional parameters $ps_i$, were included for each star, where $P$ is the number of plate-MJD pluggings composing the star.  These plate-shift parameters allow all \rvs\ from plate $i$ to shift by some amount $ps_i$, relative to the first plate $P_0$.  For the majority of stars $P=1$, no plate-shift parameters are necessary, and \smodel\ is a 1-parameter model.

  In the binary star model \bmodel, the sparsely sampled \rvs\ are fit by a sinusoid defined by four-parameters: the log of the semi-amplitude \loga, the log of the period \logp, the phase $\phi$, and the systemic velocity \vnaught.  We assume circular orbits (eccentricity, $e = 0$), which is a safe assumption for tidally circularized, short-period orbits ($P<12$ days; \citealp{raghavan10}), where we are most sensitive.  A small number of the binaries found in this study may have longer periods and could have non-zero eccentricities, but this does not affect our results.  Plate-shift parameters were also adopted in \bmodel\ wherever $P>1$.

  Uninformative priors were used in the MCMC.  The model index $\lambda$, has a flat prior from 0 to 1, where $\lambda<0.5$ denotes \smodel\ and $\lambda\geq0.5$ denotes \bmodel.  The semi-amplitude prior is log-uniform from $3\ \kmps$, comparable to the measurement uncertainties where \smodel\ and \bmodel\ become degenerate, to $250\ \kmps$, greater than the largest \rv\ differences in the sample. The prior on the period is uniform in the range $4\leq\logp\ (\second)\leq 7$. The lower limit $\logp\ (\second)=4.0$ is equal to the orbital period at which stellar contact is certain for low-mass companions.  Above $\logp\ (\second)=7.0$, \rv\ amplitudes in binary systems are comparable to the measurement uncertainties.  Combined with the sparsity of the \rv\ data, systems with periods longer than $\logp\ (\second)=7.0$ are outside our range of sensitivity.  Priors are also uniform for the phase ($0\leq\phi\leq 2\pi$) and systemic velocity ($-600\leq\vnaught\ (\kmps)\leq 600$).  Markov chains were run independently on every star with an ensemble of $200$ parallel chain ``walkers'' for a total of $2.4\e6$ samples, then burned and thinned to $6\e5$ independent samples of the posterior.

\begin{figure}[htbp]
    \centering
    \plotone{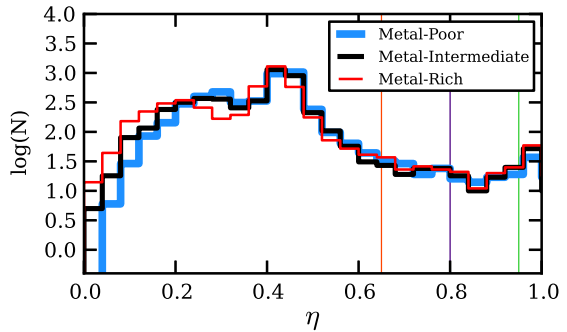}
    \caption
    {Distribution of $\eta$, the fraction of posterior samples using the binary model, for stars.}
    \label{fig:etaHist}
  \end{figure}

  Evidence for detection of a companion star is reflected by the relative probabilities of $\lambda$.  We define the probability for the binary model, $\eta$ as the fraction of samples in the marginalized posterior having $\lambda=\bmodel$.  We note that the value of $\eta$ is dependent on the choice of priors, and is sensitive to the treatment of the SDSS systematics.  Moreover, a degeneracy arises as the \rv\ curve of a  long-period, low-amplitude system becomes indistinguishable from a single-star system.  With this mind, we stress that values for $\eta$ are not absolute probabilities of a system having a companion, but reflect the ability of the data to rule out models under the given prior.  However, the \feh-groups can be compared, relatively, by considering the fraction of systems where $\eta$ is large and \smodel\ is strongly disfavored.  The results are shown in Figure \ref{fig:etaHist}

  We also investigated the \eoi\ parameter proposed by \cite{geller08} as a metric for identifying the stars with large \rv\ variations.  We find that the \eoi\ parameter singles out many of the same stars as our more sophisticated MCMC-based inference.  Our method not only takes into account deviations in \rv\ from the mean, but also how well the data fit the expected periodicity of a binary system.

  Analysis of the posterior, and visual inspections of the binary model fits, show that 681 stars with $\eta>0.65$ are true spectroscopic binaries, though given the sparsity of the RV curve sampling, there are sometimes large uncertainties in the fitted values for specific model parameters.  Another natural break point is $\eta>0.95$; these are 209 stars for which the determination and analysis of accurate individual model parameters should be possible (and will be characterized in future work).  An intermediate cut at $\eta>0.80$ is a compromise between these limits, yielding a larger sample of stars (406) with modest model constraints.  The values of the binary fractions that we derive below are insensitive, within the uncertainties, to the exact choice of cut in $\eta$.  This implies that the \rv\ variations for our binary detections are sufficiently above the measurement uncertainties, and that the binary fractions reported are not biased due to differences in \snr\ or absorption features.

\begin{figure}[htbp]
    \centering
    \plotone{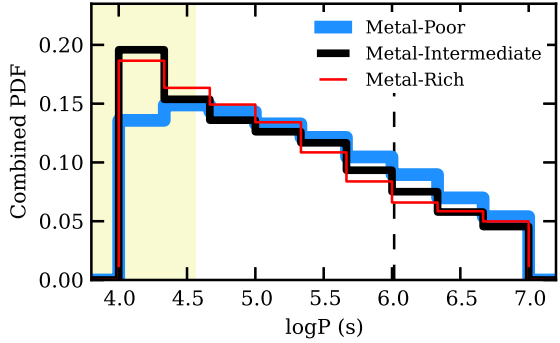}
    \caption
    {Averaged probability distributions of \logp\ for all binary detections ($\eta>0.80$).  These do not reflect actual distributions of periods, and should only be used as a guide to probe the region of MCMC sensitivity.  The shaded region indicates where Roche lobe overflow and contact becomes relevant.  The dashed line marks the circularization limit at a period of 12 days.}
    \label{fig:logP}
  \end{figure}

  Figure \ref{fig:logP} shows the \logp\ posteriors for each \feh-group, marginalized over all binary systems ($\eta>0.80$).  The posterior distributions of \logp\ for many of these stars are complex: many are multimodal, affected by aliasing or other issues related to the sparse, biased time sampling.  One such effect is the increase in probability at $\logp=4$. Here the metal-rich and -intermediate groups contain more stars than the metal-poor group with $\Delta t\simeq10^4\ \second$.  Systems with periods as short as this are extremely rare \citep{drake14}, and our increased probability in this area may be due to overfitting. Additionally, the gap at $\Delta t=10^{4.6}\ \second=12\ \hour$ (Figure \ref{fig:feh_baseline}) may affect the estimate of a period.  We defer a more sophisticated analysis to a future paper, but these effects should not alter the ability to rule out a single-star model.  For now, Figure \ref{fig:logP} illustrates that we are mainly sensitive to periods in the range $4<\logp\ (\second)<6$, or less than about 12 days.  We emphasize that a more detailed analysis will be necessary to estimate the true underlying \logp\ distribution in our sample.

  \begin{figure*}[htbp]
    \centering
    \plotone{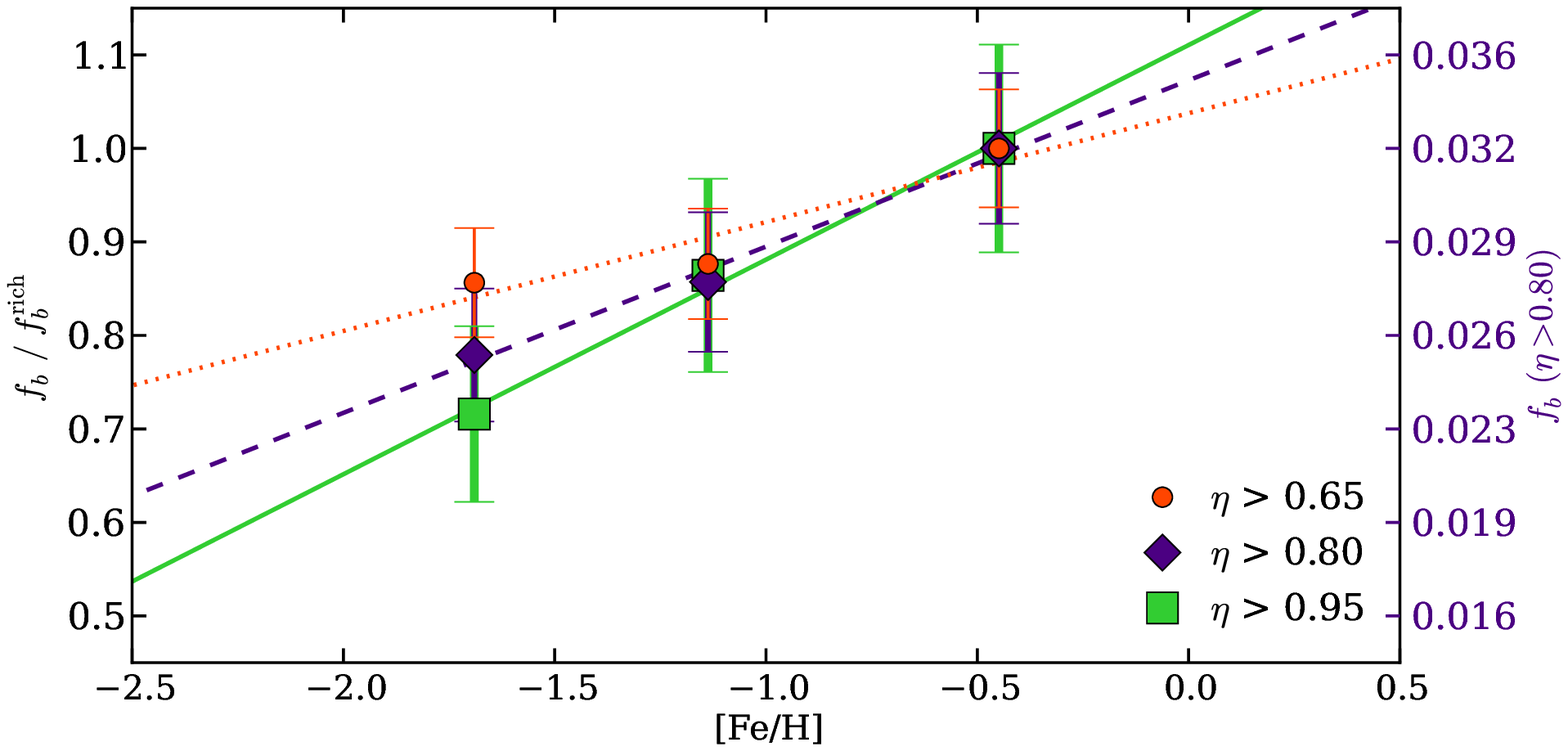}
    \caption
    {Short-period binary fraction limits, relative to the metal-rich group.  Binary companion detections are defined by a cut in $\eta$, the fraction of posterior samples using the binary model.  Group median values of \feh\ are used.}
    \label{fig:binaryFractions}
  \end{figure*}

\section{Discussion}
  \label{sec:discussion}
  In Figure \ref{fig:binaryFractions} we show \fbin, the measured lower bound for the fraction of stars with short-period companions $(P\lesssim 12\ \mathrm{days})$ for each metallicity group, normalized to the metal-rich binary fraction.  \fbin\ is a lower limit because of non-detections as a result of sparsely sampled \rvs\ and high orbital inclinations, resulting in low amplitudes.  We see agreement in \fbin\ measured for all three choices in $\eta$ cutoff (0.65, 0.80, 0.95).  With a cutoff of $\eta=0.80$, values of \fbin\ for the metal-poor, -intermediate, and -rich groups respectively are: $2.5\%\pm0.2\%$, $2.8\%\pm0.2\%$, and $3.2\%\pm0.3\%$.  Since the observational biases that affect binary detection are mostly due to the sparsity of the RV coverage, which isn't metallicity-dependent, we conclude that the field F-type MS stars in our metal-rich sample are, at a 2-sigma level, $30\%$ more likely than those in our metal-poor sample to have close binary companions.  

  Our metal-rich and metal-poor samples mostly trace the MW disk and halo. Differences in the fraction of short-period systems can stem from differences in the star-formation process, dynamical interactions after star formation, or some combination of the two.

  Three-dimensional hydrodynamic models from \cite{machida09} actually suggest a \textit{higher} frequency of binaries formed through cloud fragmentation for metal-poor clusters, due to the decreased requirement of a cloud's initial rotation energy to fragment.  Moreover, their models yield systems with shorter initial separations at lower metallicities.  The increased \fbin\ observed for metal-rich stars in this work can more likely be explained by dynamical processes than by formation processes.  

  The observed differences in \fbin\ could be explained if the clusters that yielded halo field stars had larger stellar densities and/or gas densities than those of the disk.  \cite{korntreff12} explore the effects of gas-induced orbital decay on period distributions in clusters.  They note that an increased density of gas in a newly formed cluster will lead to a larger number of short-period system mergers shortly after formation.  \cite{parker09} describe how clusters with higher stellar densities destroy wide binaries through dynamical interactions.  An increase in the destruction of high-mass, wide-binary systems leads to the ejection of former F-star secondaries into the field.  These orphaned, single-star systems would increase the total number of F-star systems in the halo field, effectively decreasing the short-period binary fraction measured.  Observational evidence of these denser cluster environments is needed to support these arguments for a lower \fbin\ in the halo.

  Additionally, some close binaries may also transfer mass and covert themselves into blue stragglers \citep{lu10}.  Evidence for an abundance of blue stragglers in the halo has been seen \citep{yanny00}, and may contribute to the lower \fbin\ observed in the metal-poor group.  Also, \cite{duchene13} show a decrease in \fbin\ with age for Solar-type stars, although this result is based on visual binaries with wider periods, and is poorly constrained due to limited sample sizes.
 
  We note that the recent results of \cite{gao14} and \cite{yuan14}, using data from SDSS,  show a larger binary fraction for metal-poor than metal-rich FGK stars in the field.  In addition to probing longer periods, the former work does not make use of sub-exposure information (using only two \rv\ epochs per star) and relies on the correctness of model values for the period distribution, mass ratio distribution, and initial mass function.  The latter work, which uses photometric color deviations to infer companions, shows a modest metallicity dependence on total binary fraction.  Since their method is not sensitive to period, the binary fractions they report are strongly dominated by more common, wider-period systems near the peak of a log-normal period distribution ($\overline{\logp}\ (\second)=10$ for nearby, Solar-like stars; \citealt{raghavan10}).  It is clear that conclusions about binary fraction depend on a number of factors, especially the range of periods to which the search is sensitive and assumptions made about the overall period distribution.

  Our MCMC analysis yields posterior probabilities in parameter space, allowing for a more detailed study of binary properties (e.g., period and separation distributions), which will be presented in future work. The techniques in this work have direct applications for current and future multiplexed spectroscopic surveys.

  \acknowledgments We thank Ewan Cameron, Dan Maoz, Jeffrey Newman, Chad Schafer, and the referee for useful discussions.  T.H. and T.C.B. acknowledge partial support from grants PHY 08-22648; Physics Frontier Center/JINA, and PHY 14-30152; Physics Frontier Center/JINA Center for the Evolution of the Elements (JINA-CEE), awarded by the US National Science Foundation.  Funding for SDSS-III has been provided by the Alfred P. Sloan Foundation, the Participating Institutions, the National Science Foundation, and the U.S. Department of Energy Office of Science.

\pagebreak
\bibliographystyle{apj}

\end{document}